\documentclass[aps,pra,twocolumn,superscriptaddress,amssymb,showpacs]{revtex4}

\usepackage{graphicx}

\begin{document}


\title{Quantum interference and sub-Poissonian statistics for time-modulated driven dissipative nonlinear oscillator}



\author{T.~V.~Gevorgyan}
\email[]{t_gevorgyan@ysu.am}
\affiliation{Institute for Physical Research, National Academy of
Sciences,\\Ashtarak-2, 0203, Armenia}

\author{A.~R.~Shahinyan}
\email[]{anna_shahinyan@ysu.am}
\affiliation{Yerevan State University, A. Manoogian 1, 0025,
Yerevan, Armenia}

\author{G.~Yu.~Kryuchkyan}
\email[]{kryuchkyan@.ysu.am}
\affiliation{Institute for Physical Research, National Academy of
Sciences,\\Ashtarak-2, 0203, Armenia} \affiliation{Yerevan State
University, A. Manoogian 1, 0025, Yerevan, Armenia}


\begin{abstract}
We show that quantum-interference phenomena can be realized for
the dissipative nonlinear systems exhibiting hysteresis-cycle
behavior and quantum chaos. Such results are obtained for a driven
dissipative nonlinear oscillator with time-dependent parameters
and take place for the regimes of long time intervals exceeding
dissipation time and for macroscopic levels of oscillatory
excitation numbers. Two schemas of time modulation: (i) periodic
variation of the strength of the ${\chi}(3)$ nonlinearity; (ii)
periodic modulation of the amplitude of the driving force, are
considered. These effects are obtained within the framework of
phase-space quantum distributions. It is demonstrated that the
Wigner functions of oscillatory mode in both bistable and chaotic
regimes acquire negative values and interference patterns in parts
of phase-space due to appropriately time-modulation of the
oscillatory nonlinear dynamics. It is also shown that the
time-modulation of the oscillatory parameters essentially improves
the degree of sub-Poissonian statistics of excitation numbers.
\end{abstract}


\maketitle

\section{\label{sec:intro}INTRODUCTION}
In recent years the study of quantum dynamics of oscillators with
time-dependent parameters has been focus of considerable
attention. This interest is justified by many applications in
different contexts. Particulary, one application concerns to the
center of mass motion of a laser cooled and trapped ion in a Paul
trap \cite{trapped_ion}. The quantum dynamics of an anharmonic
oscillator (AHO) with time dependent modulation of its frequency
and nonlinearity parameters has been investigated in applications
to macroscopic superposition of quantum states
\cite{time_dep_nonlinearity}. In the last few years there has been
rapid progress in the construction and manipulation of
nanomechanical oscillators with giant ${\chi}(3)$-Kerr
nonlinearity \cite{nano}-\cite{nano_2}. The nanomechanical
resonator with a significant fourth-order nonlinearity in the
elastic potential energy has been experimentally demonstrated
\cite{A}. It has also been shown that this system is dynamically
equivalent to the Duffing oscillator with varied driving force
\cite{duffing_force}. This scheme is widely employed for a large
variety of applications as well as the other schemes of micro- and
nanomechanical oscillators, more commonly as sensors or actuators
in integrated electrical, optical, and optoelectrical systems
\cite{nano}, \cite{devices}.

It is well assessed that in the case of unitary dynamics, without
any losses, an anharmonic oscillator leads to sub-Poissonian
statistics of oscillatory excitation number, quadratic squeezing
and superposition of macroscopically distinguishable coherent
states. For dissipative dynamics the important parameter
responsible for production of nonclassical states via ${\chi}(3)$
materials is the ratio between nonlinearity and damping.
Therefore, the practical realization of such quantum effects
requires a high nonlinearity with respect to dissipation. In this
direction the largest nonlinear interaction was proposed in many
papers, particularly, in terms of electromagnetically induced
transparency \cite{EM} and by using the Purcell effect \cite{P},
and in cavity QED \cite{R}. The significant nonlinearity has also
been observed for nanomechanical resonators \cite{N}. These
methods can lead to ${\chi}(3)$ nonlinearity of several orders of
magnitude higher than natural optical self-Kerr interactions.
Note, that high ${\chi}(3)$ nonlinear oscillators generate also a
lot of interest recently due to their applications in areas of
quantum computing \cite{C}.

In the case of nonlinear dissipative ${\chi}(3)$ interaction
stimulated by coherent driving force, the time evolution cannot be
solved analytically for arbitrary evolution times and suitable
numerical methods have to be used. Nevertheless, with dissipation
included a driven AHO model has been solved exactly in the
steady-state regime in terms of the Fokker-Planck equation in
complex P representation \cite{I}. Analogous solution has been
obtained for a combined driven parametric oscillator with Kerr
nonlinearity \cite{II}. The Wigner functions for both these models
have been obtained using these solutions \cite{I}, \cite{III}.

 The investigation of quantum dynamics of a driven dissipative
nonlinear oscillator for non-stationary cases is much more
complicated and only a few papers have been done in this field up
to now. More recently, the quantum version of dissipative AHO or the
Duffing oscillator with time-modulated driving force has been
studied in the series of the papers \cite{SR}, \cite{X1},
\cite{X2} in the context of a stochastic resonance \cite{SR},
quantum-to-classical transition and investigation of quantum
dissipative chaos \cite{X1}, \cite{X2}.

 In this paper we continue investigation of time-modulated effects
for quantum version of the Duffing oscillator. In this direction
we investigate quantum effects in the presence of dissipation and
decoherence, mainly sub-Poissonian statistics and signature of
quantum interference within the framework of the Wigner function
for an oscillatory mode. Two schemas of time modulation: (i)
periodic variation of the strength of the ${\chi}(3)$
nonlinearity; (ii) modulation of the amplitude of the driving
force will be considered. Our main result is that time modulation
of the oscillatory parameters essentially improves the degree of
sub-Poissonian statistics of oscillatory excitation numbers as
well as leads to negative values of the Wigner functions in an
over transient regime for definite time intervals exceeding the
transient time within the period of modulation. Thus, we
demonstrate that for the case of time-modulated AHO the Wigner
function gradually deviates from the Wigner function corresponding
to an ordinary AHO without any modulation. Really, surprisingly
simple analytical results for the Wigner functions of an ordinary
AHO have been analytically obtained in over transient regime
\cite{I}, \cite {III}, that are positive in all ranges of
phase-space. In our scheme the negative values for the Wigner
function of oscillatory mode, that reflect quantum-interference
patterns in phase space, appear due to time-modulation dynamics
and are realized in both bistable and chaotic operational regimes.

Note, that nowadays, large theoretical and experimental papers
have been devoted to the implementation of macroscopic (i.e., many
particle) quantum superposition states (MQS). The most notable
results on MQS are experimentally obtained with atoms interacting
with microwave fields trapped inside a cavity \cite{SC1} or for
freely propagating fields \cite{SC2} and with optical parametric
amplifier in the presence of decoherence \cite{SC3}. Naturally,
the question arises whether the quantum-interference patterns
obtained for driven AHO can be realized for time intervals
exceeding the characteristic decoherence and dissipation time on
many particle level. The corresponding results obtained here are
verified for the operational regimes of high nonlinearity with
respect to dissipation, e.g., for the ratio ${\chi}/{\gamma} \simeq
1\cdot 10^{-1}\div 1.5\cdot 10^{-3}$, where $\chi$ is the
nonlinear constant proportional to ${\chi}(3)$ and $\gamma$ is the
damping constant. Nevertheless, we have checked for these
parameters that phase-space interference patterns are also
obtained in the macroscopic level that involves a number of
oscillatory excitation numbers in excess of $52$.

 The outline of this paper is as follows. In the next section we
describe the models and investigate the mean oscillatory
excitation numbers and quantum fluctuations of excitation numbers
on the base of the Mandel parameters. In Sec. III we shortly
discuss the case of the unitary dynamics. In Sec. IV we present
the results for the Wigner functions showing negativity due to the
time modulations of the oscillatory dynamics. We summarize our
results in Sec. V.

\section{\label{sec:model}Models: excitation numbers and quantum statistics}
We treat the Duffing oscillator as an open quantum system and
assume that its time evolution is described by Markovian dynamics
in terms of the Lindblad master equation for the reduced density
matrix $\rho$. In the interaction picture that corresponds to the
transformation $\rho\longrightarrow e^{-i\omega a^{+}at} \rho
e^{i\omega a^{+}at}$, where $a^{+}$ and $a$ are the Bose
annihilation and creation operators of the oscillatory mode and
$\omega$ is the driving frequency, this equation reads in the
Markovian form as
\begin {eqnarray}
\frac{d\rho}{dt}= \frac{-i}{\hbar}[H_0+H_{int}, \rho]+\nonumber~~~~~~~~~~~~~~~~~~~~~~~~~~~~~~~~\\
\sum_{i=1,2}\left( L_{i}\rho
L_{i}^{+}-\frac{1}{2}L_{i}^{+}L_{i}\rho-\frac{1}{2}\rho L_{i}^{+}
L_{i}\right).\label{eq:master_equatin}
\end {eqnarray}
The Hamiltonians are
\begin{eqnarray}
H_{0}=\hbar\Delta a^{+}a, \nonumber ~~~~~~~~~~~~~~~~~~~~~~~~~~~~~~~~~~~~\\
H_{int}=\hbar\chi(t) (a^{+}a)^{2}+\hbar(f(t)a^{+}+
f(t)^{*}a)\label{eq:peturbation},
\end{eqnarray}
where $\chi(t)$ and $f(t)$, which may or may not depend on time,
represent, respectively, the strength of the nonlinearity and
amplitude of the force, $\omega_0$ is the resonant frequency,
$\Delta=\omega_0-\omega$ is the detuning. The dissipative and
decoherence effects, losses, and thermal noise are included in the
last part of this equation, where $L_{i}$ are the Lindblad
operators:
\begin {equation} \label{eq:linbland}
L_{1}=\sqrt{(N+1)\gamma}a, \qquad L_{2}=\sqrt{N\gamma}a^{+},
\end{equation}
 $\gamma$ is the spontaneous decay rate of the dissipation
process and $N$ denotes the mean number of quanta of a heat bath.

This model seems experimentally feasible and can be realized in
several experimental schemes. In fact, a single mode field is well
described in terms of an AHO, and the nonlinear medium could be an
optical fiber or a $\chi(3)$ crystal, placed in a cavity. The
anharmonicity of mode dynamics comes from the self-phase
modulation due to the photon-photon interaction in the $\chi(3)$
medium. In this case, it is possible to realize time modulation of
the strength of the nonlinearity by using a media with periodic
variation of the $\chi(3)$ susceptibility.

On the other side, the Hamiltonian described by Eq.
(\ref{eq:peturbation}) describes a single nanomechanical resonator
with $a^{+}$ and $a$ raising and lowering operators related to the
position and momentum operators of a mode quantum motion
\begin {equation}
x=\sqrt{\frac{\hbar}{2m\omega_0}}(a+a^{+}),\nonumber \
p=-i\sqrt{2\hbar m\omega_0}(a-a^{+}) \label{eq:xp},
\end{equation}
where $m$ is the effective mass of the nanomechanical resonator,
$\omega_0$ is the linear resonator frequency and $\chi$
proportional to the Duffing nonlinearity. One of the variants of
nano-oscillators is based on a double-clamped platinum beam
\cite{N} for which the nonlinearity parameter equals to
$\chi=\hbar/4\sqrt{3}Qma_{c}^{2}$, where $a_c$ is the critical
amplitude at which the resonance amplitude has an infinite slope
as a function of the driving frequency, $Q$ is the mechanical
quality factor of the resonator. In this case, the giant
nonlinearity $\chi\cong3.4\cdot10^{-4} s^{-1}$ was realized. Note,
that details of this resonator, including expression for the
parameter $a_{c}$, are presented in \cite{amp}. On decreasing
nanomechanical resonator mass, its resonance frequency increases,
exceeding 1 GHz in recent experiments \cite {nano}, \cite
{nano_1}. It is possible to reach a quantum regime for such
frequencies, i.e., to cool down the temperatures for which thermal
energy will be comparable to the energy of oscillatory quanta. The
recent investigations in this direction are devoted to classical
to quantum transition of a driven nanomechanical oscillator
\cite{Qnr1}, generation of Fock states \cite{Qnr2}, nonlinear
dynamics, and stochastic resonance \cite{Qnr3}.

Cyclotron oscillations of a single electron in a Penning trap with
a magnetic field are another realization of the quantum version of
the Duffing oscillator \cite{a}, \cite{b}, \cite{c}. In this case
the anharmonicity comes from nonlinear effect that is caused by
the relativistic motion of an electron in a trap, while the
dissipation effects arise from the spontaneous emission of the
synchrotron radiation and thermal fluctuations of the cyclotron
motion. Note that a one-electron oscillator allows one to achieve
a relatively strong cubic nonlinearity, $\chi/\gamma \lesssim 1$.

 For the constant parameters $\chi(t)=\chi$ and $f(t)=f$ the equations
(\ref{eq:master_equatin}) and (\ref{eq:peturbation}) describe the
model of a driven dissipative AHO that was introduced long ago in
quantum optics to describe bistability due to a Kerr nonlinear
medium \cite{drumm}. For the case of time-dependent parameters
$\chi(t)$ and $f(t)$ the dynamics of the AHO exhibits a rich
phase-space structure, including regimes of regular, bistable and
chaotic motion. We perform our calculations for these three
qualitatively different regimes concerning two models of
time-modulated AHO corresponding to two physical situations: (i)
$\chi=\chi(t)=\chi_0 + \chi_1 sin(\delta t)$ and $f(t)=const=f$;
(ii) $f=f(t)=f_0 + f_1 sin(\Omega t)$ and $\chi(t)=const=\chi$
with $\delta \ll \omega$ and $\Omega \ll\omega$ the modulation
frequencies.

We analyze the master equation numerically using quantum state
diffusion method (QSD) \cite{QSD}. According to this method, the
reduced density operator is calculated as the ensemble mean
\begin{equation}
\rho(t)= M(|\psi_{\xi}(t)\rangle\langle\psi_{\xi}(t)|)= \lim_
{N\rightarrow\infty}\frac{1}{N}\sum_{\xi}^{N}|\psi_{\xi}(t)\rangle\langle\psi_{\xi}(t)|
\end{equation}
over the stochastic pure states $|\psi_{\xi}(t)\rangle$ describing
evolution along a quantum trajectory. The stochastic equation for
the state $|\psi_{\xi}(t)\rangle$ involves both Hamiltonian
described by Eq. (\ref{eq:peturbation}) and the Linblad operators
described by Eq. (\ref{eq:linbland}). We calculate the density
operator using an expansion of the state vector
$|\psi_{\xi}\rangle$ in a truncated basis of Fock's number states
of a harmonic oscillator
\begin{equation}
|\psi_{\xi}(t)\rangle= \sum_{n}a_{n}^{\xi}(t)|n\rangle.
\end{equation}

\subsection{\label{sec:sub_pois}Improvement of the sub-Poissonian statistics by modulation of the nonlinearity}

In the classical limit the corresponding equation of motion for
the dimensionless mean amplitude $\alpha=Tr(\rho a(t))$ has the
form
\begin{equation}
\frac{d\alpha}{dt}=-\frac{\gamma}{2}\alpha-i(\Delta
+(\chi_0+\chi_1\sin\delta
t)(1+2|\alpha|^{2}))\alpha-if.\label{eq:alpha}
\end{equation}
We remind the reader that for the constant strength of the
nonlinearity, $\chi_1=0$, the semiclassical mean excitation number
$|\alpha|^2$ as a function of the driving intensity can exhibit
multiple steady states and hysteresis. In the case of modulation,
$\chi_1\neq0$, the interaction Hamiltonian is explicitly time
dependent, and the system exhibits regions of regular, bistable,
and chaotic motion with $\chi_0$, $\chi_1$, $\Delta$, and the
modulation frequency being the control parameters.

We examine these operational regimes by numerical analysis of the
phase space of dimensionless position and momentum $x=Re(\alpha)$
and $y=Im(\alpha)$. Choosing $x_0$ and $y_0$ as an arbitrary
initial phase-space point of the system at the time $t_0$, we
define a constant phase map in the $(X, Y)$ plane by the sequence
of points $(X_n, Y_n)=(X(t_n), Y(t_n))$ at
$t_n=t_0+\frac{2\pi}{\delta}n$ ($n=0, 1, 2, ...$). This means that
for any $t=t_n$ the system is at one of the points of the
Poincar\'{e} section. The analyses show that for time scales
exceeding the damping time $1/\gamma$, the asymptotic dynamics of
the system is regular in the limits of small and large values of
the modulation frequency, i.e., $\delta \ll \gamma $, $\delta
\gg\gamma $, and for positive values of the detuning, $\Delta >
0$. Such regime is also realized when the modulation part of the
parameter, i.e., $\chi_{1}$ is much less than the $\chi_{0}$.
\begin{figure}
\includegraphics[width=8.6cm]{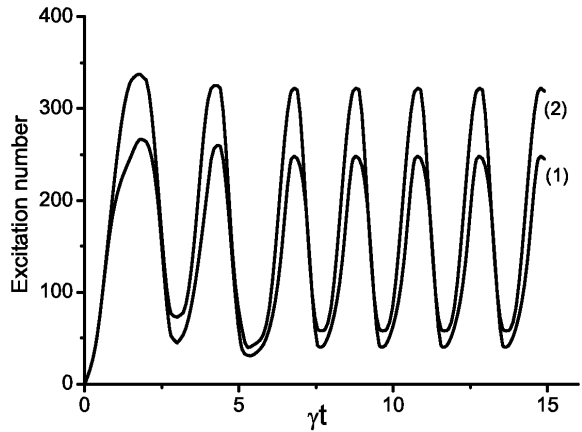}
\caption{The photon excitation numbers for two cases: (1) for
monostable regime with parameters $\Delta/\gamma=0.1,
\chi(t)/\gamma=5\cdot10^{-3}(1+0.7\sin(\delta t)),
\delta/\gamma=3, f/\gamma=20$; (2) for bistable regime with
parameters $\Delta/\gamma=-1,
\chi(t)/\gamma=6\cdot10^{-3}(1+0.5\sin(\delta t)),
\delta/\gamma=3, f/\gamma=20$.}\label{fig:number}
\end{figure}

First, we discuss the case of classically regular behavior
assuming the interaction of the system with vacuum reservoir:
$N=0$. Below we present numerical results for both the oscillatory
mean excitation number $\langle
n\rangle=M(\langle\psi_{\xi}|a^{+}a|\psi_{\xi}\rangle)$ and Mandel
$Q$ parameter which describes the deviation of excitation number
uncertainty from the Poissonian variance, i.e.,
$Q=\frac{\langle(\Delta n)^{2}\rangle-\langle
n\rangle}{\langle n\rangle }$, $\langle(\Delta
n)^{2}\rangle=\langle (a^{+}a)^{2}\rangle-\langle
a^{+}a\rangle^{2}$, i.e.,
\begin{equation}
\langle(\Delta n)^{2}\rangle=M(\langle
\psi_{\xi}|(a^{+}a)^{2}|\psi_{\xi}\rangle)-\langle n\rangle^{2}.
\end{equation}
\begin{figure}
\includegraphics[width=8.6cm]{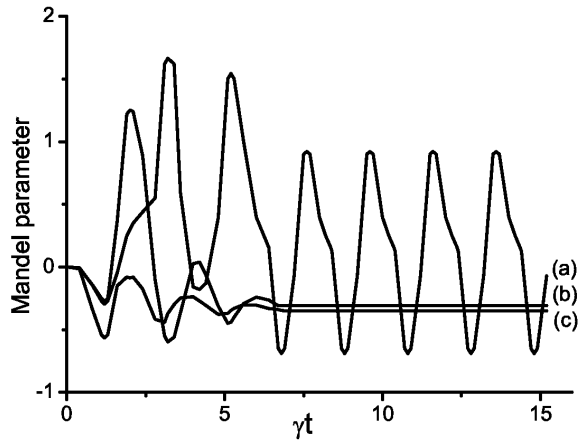}
\caption{The Mandel parameter for the monostable regime with the
parameters as in Fig. \ref{fig:number}, case (1), (curve(a)); AHO without
time-modulation:
 $\chi/\gamma= 8.5\cdot10^{-3}$ (b); $\chi/\gamma=1.5\cdot10^{-3}$ (c). The other parameters
are as in the Fig. \ref{fig:number}, case (1).}\label{fig:mandel}
\end{figure}
For the driven dissipative AHO under time-modulation of the
strength of nonlinearity, the ensemble-averaged mean oscillatory
excitation number exhibits a periodic time-dependent behavior in
both cases of regular and chaotic regimes for over transient
time-intervals. For bistable and monostable regimes the results of
numerical calculations for time evolution of $\langle n \rangle$
are depicted in Fig. \ref{fig:number}(case 1) and Fig. \ref{fig:number}(case 2) for the
parameters: $\chi(t)/\gamma=5\cdot10^{-3}(1 + 0.7\sin(3 \gamma
t))$ and $\chi(t)/\gamma=6\cdot10^{-3}(1+0.5\sin(3 \gamma t))$.
The time evolution of the Mandel parameter is depicted on Fig. \ref{fig:mandel}.
As we see (Fig. \ref{fig:mandel}(a)), the $Q$ parameter also shows a
time-dependent modulation and formation of the sub-Poissonian
statistics $(Q < 0)$ for the definite time intervals exceeding the
transient time $t\gamma \geq 6$. The level of the oscillatory
excitation-number fluctuations reaches to the minimum values
$Q_{min}=-0.69$ and the maximum values $Q_{max}=1.2$ at the fixed
intervals of time, respectively, at $\gamma t_{k}=6.8 +\frac{2\pi
k}{\delta}\gamma$ and at $\gamma t_{k}=7.6+ \frac{2\pi
k}{\delta}\gamma$ , $(k=0,1,2...)$. Comparing this result with the
case without any time-modulation we also present the $Q$ parameter
for the maximal and minimal values of $\chi(t)$, i.e., for
$\chi=8.5\cdot10^{-3}$ ($Q=-0.3$, curve (b)) and for
$\chi=1.5\cdot10^{-3}$ ($Q=-0.36$, curve (c)). Thus, we conclude
that the modulation of the strength of nonlinearity leads to the
improvement of the level of sub-Poissonian statistics below the
level of corresponding steady-state regime. Indeed, we find that
$Q_{min}<-0.36$. It is interesting that $Q_{min}$ corresponds to
the maximal value of the mean excitation number, $\langle
n\rangle=249$. As calculations show, analogous result takes place
for the bistable regime, Fig. \ref{fig:number}, case(2). In this case,
$Q_{min}=-0.7$ for the mean excitation numbers $\langle n\rangle=
352 $. Note that such effect of improving the sub-Poissonian
statistics has been recently obtained for the dissipative AHO
under time-modulation of the driving amplitude \cite{X1}.

\section{\label{sec:unit_dyn}Time modulation in the case of unitary dynamic}

We now calculate the quantum distributions, at first considering
the simplest case of non-dissipative AHO without any driving. In
this case the operator $(a^{+}a)$ is an invariant of motion, since
this operator commutes with the corresponding Hamiltonian
$H(t)=\hbar[\Delta a^{+}a + \chi(t)(a^{+}a)^{2}]$. The system
exhibits an unitary evolution; for an initial coherent state
$|\alpha_{0}\rangle$ with the phase $\varphi=0$ at $t=0$, the
state evolution at time $t$ is
\begin{equation}
|\psi(t)\rangle=U\left(t\right)|\alpha_{0}\rangle=e^{-|\alpha_{0}|/2}\sum_{n=0}^{\infty}\frac{|\alpha_{0}|^{n}}{\sqrt{n!}}
e^{-i\varphi(t)n^{2}}|n\rangle \label{eq:vec},
\end{equation}, where $\varphi(t)=\int_{0}^{t}\chi(\tau )d \tau$. This state involves the time-modulation effects as
well as describes the interference of the Fock number states $|n
\rangle$. The corresponding nonstationary Wigner function is
written as
\begin{equation} \label{eq: Wigner_alpha}
W(\alpha, t)=\frac{2}{\pi^{2}}e^{-2|\alpha|^{2}}\int d^2\beta
\langle -\beta|\psi(t)\rangle\langle \psi(t)|\beta\rangle
e^{-2(\beta\alpha^{*}-\beta^{*}\alpha)}.
\end{equation}
We transform this formula to the form
\begin{equation}\label{eq:WigRT}
W(r, \theta)=\sum_{n,m}\rho_{nm}(t)W_{mn}(r,\theta),
\end{equation}
where $(r,\theta)$ are the polar coordinates in the complex
phase-space plane, $x=rcos\theta$, $y=rsin\theta$, while the
coefficients $W_{mn}(r,\theta)$ are the Fourier transform of
matrix elements of the Wigner characteristic function

\begin{widetext}
\begin{eqnarray}\label{eq:Wigf}
W_{mn}\left(r,\theta \right) = \left\{
\begin{array}{rcl}
\frac{2}{\pi}\left(-1\right)^{n}\sqrt{\frac{n!}{m!}}e^{i(m-n)\theta}(2r)^{m-n}e^{-2r^2}L_{n}^{m-n}(4r^2),~ m\geq n\\
\frac{2}{\pi}\left(-1\right)^{m}\sqrt{\frac{m!}{n!}}e^{i(m-n)\theta}(2r)^{n-m}e^{-2r^2}L_{m}^{n-m}(4r^2),~ n\geq m
\end{array},
\right.
\label{FirstOrderNPlusSQRSolution}
\end{eqnarray}
\end{widetext}

The matrix elements $\rho_{nm}(t)=\langle n|\psi(t)\rangle \langle
\psi(t)|n\rangle$ are equal to
\begin{equation}\label{eq:W13}
\rho_{nm}(t)=e^{-|\alpha_{0}|^{2}}\frac{|\alpha_{0}|^{n+m}}{\sqrt{n!m!}}
e^{i\varphi(t)(n^{2}-m^{2})},
\end{equation}
where for $\chi(t)=\chi_{0}+\chi_{1}sin(\delta t+\phi)$ we have
\begin{equation}
\varphi(t)=\chi_{0} t + \frac{\chi_{1}}{\delta}[\cos(\delta t +
\phi)- \cos\phi].
\end{equation}
These results are valid for short time intervals, much less than
the characteristic dissipation time, $t\ll\gamma^{-1}$. The Wigner
function Eq. (\ref{eq:WigRT}) with the density matrix Eq.
(\ref{eq:W13}) leads to the interference fringes arising from the
non-diagonal elements $\rho_{nm}$. For non-modulated case,
$\varphi(t)=\chi_{0} t$, the Wigner function is well studied (see,
for example, \cite{cccc}). Particularly, for $\chi
t=\frac{\pi}{2}$ and in the case of the initial coherent state, it
describes a superposition of other coherent states of the same
amplitude but different phases. The novelty of the Eqs.
(\ref{eq:WigRT}) and (\ref{eq:W13}) consists in the time-dependent
nonlinearity that is the factor of $\varphi(t)$. This modulation
term allows to control the time intervals of the maximal
superposition with appropriate choice of the modulation frequency
and the phase $\phi$. Indeed, the first time at which we have the
superposition of coherent states is determined from
$\varphi(t)=\frac{\pi}{2}$. Particularly, for $\delta t\ll 1$, we
have $\varphi(t)\simeq(\chi_{0}+\chi_{1}\sin\phi)t$ and hence the
superposition is realized, if
$t=\pi/2(\chi_{0}+\chi_{1}\sin\phi)$.

\section{\label{sec: quan_inter_patterns}Wigner functions and quantum-interference patterns}

It is well known that the phase-space Wigner distribution
function can simply visualize nonclassical effects including
quantum-interference. For example, a signature of quantum
interference is exhibited in the Wigner function by the
non-positive values. In this section the numerical results of the
nonstationary Wigner functions in bistable and chaotic regimes of
AHO are presented and discussed.

It should be noted that the most of investigations of the quantum
distributions of oscillatory states, including also modes of
radiations, have been made for the steady-state situations. The
simplicity of Kerr nonlinearity allows to determine the Wigner
function of the quantum state under time evolution due to
interaction (see, for example \cite{cccc}). In this sense, we note the main
peculiarity of our paper in comparison with above noted important
inputs. In this paper, we calculate the Wigner functions in an over
transient regime, $t\gg\gamma^{-1}$, of the dissipative dynamics,
however, we consider time-dependent effects which appear due
to the time-modulation of the oscillatory parameters.

Below we investigate the Wigner functions for both cases of
time-modulation (see, cases (i) and (ii)). For periodic variation
of the strength of $\chi(3)$ nonlinearity, semiclassical dynamics
is described by Eq. (\ref{eq:alpha}), while for the case of
time-modulation of the driving amplitude the semiclassical
equation reads as
\begin{equation}\label{eq:alpha_force}
\frac{d\alpha}{dt}=-\frac{\gamma}{2}\alpha-i(\Delta
+\chi_0(1+2|\alpha|^{2}))\alpha-i(f_0+f_1 sin(\Omega t)).
\end{equation}
\begin{figure}
\includegraphics[width=8.6cm]{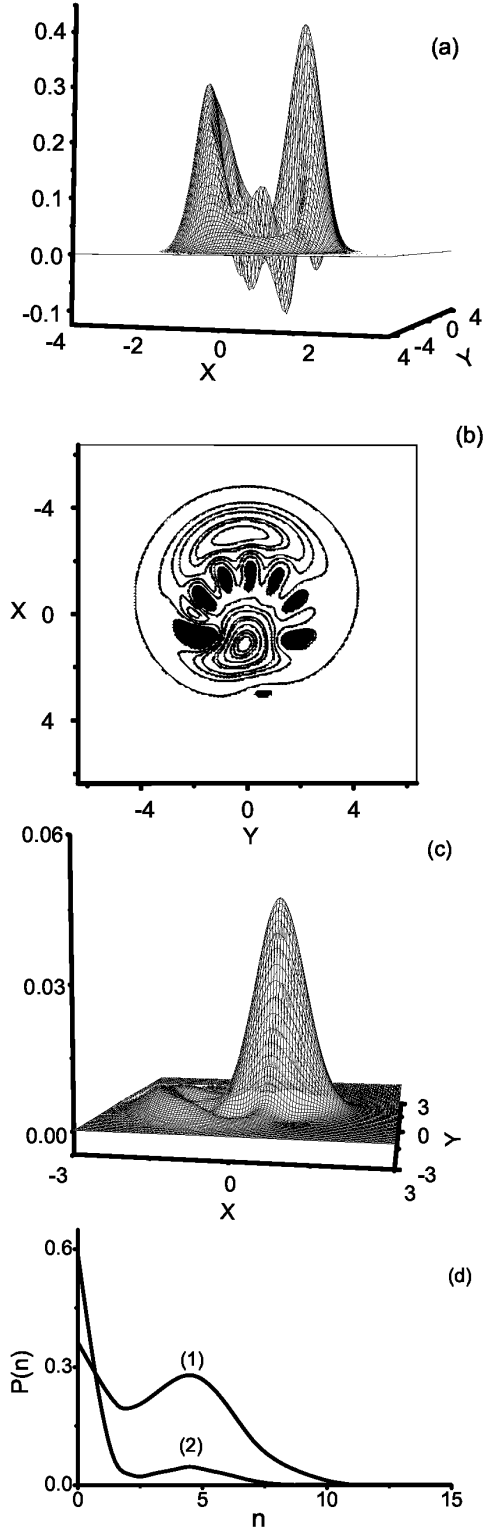}
\caption{ The Wigner function and its contour plot, averaged over
3000 trajectories, for the parameters: $\Delta/\gamma=-15,
\chi/\gamma=2, f/\gamma=5.8(1 + 0.5\sin(\Omega t)),
\Omega/\gamma=2$, (a) and (b). The contour plot of the Wigner
function that indicates the ranges of negativity in the black (b).
The Wigner function for AHO without modulation, $f/\gamma=5.8$
(c). The probability distribution of excitation numbers (d): case
of time-modulation (1), case without of modulation (2). The
parameters are as on the Figs. (a) and (c), respectively.}\label{fig:bi}
\end{figure}
\begin{figure}
\includegraphics[width=8.6cm]{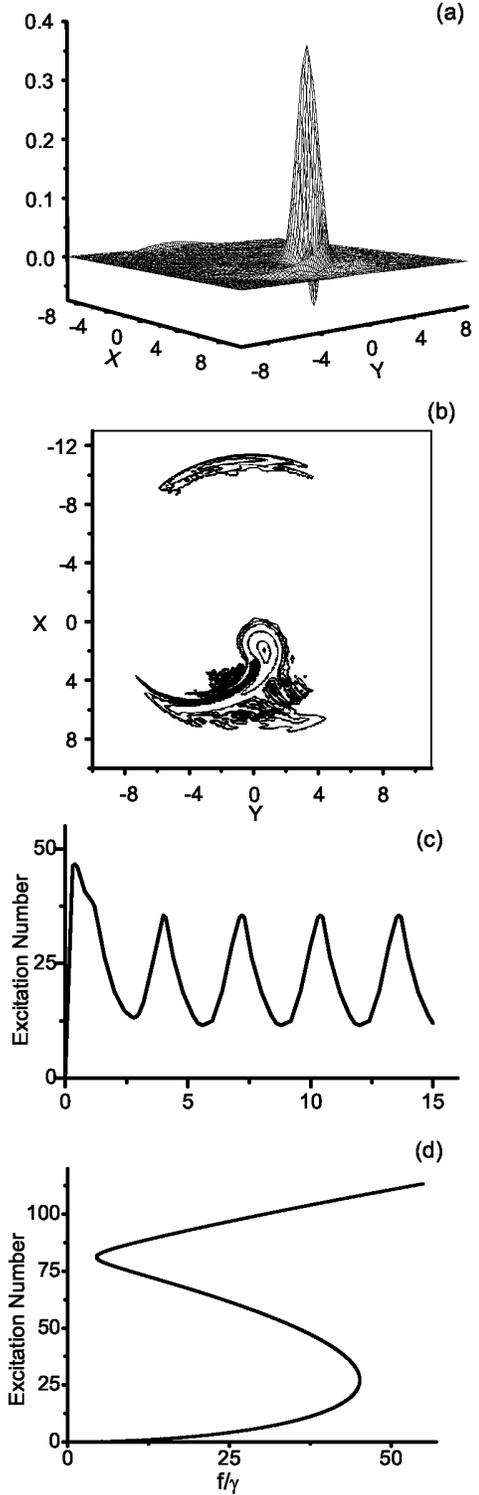}
\caption{The Wigner function (a); the contour plot of the Wigner
function (b); and the mean excitation number (c); averaged over
3000 trajectories, for the dissipative AHO under time-modulated
driving. The parameters are in the range of bistability:
$\Delta/\gamma=-13.02, \chi/\gamma=0.08, f/\gamma=29(1 +
0.5\sin(\Omega t)), \Omega/\gamma=2$. The hysteresis curve (d) for
the stationary case.}\label{fig:scaled}
\end{figure}
\begin{figure}
\includegraphics[width=8.6cm]{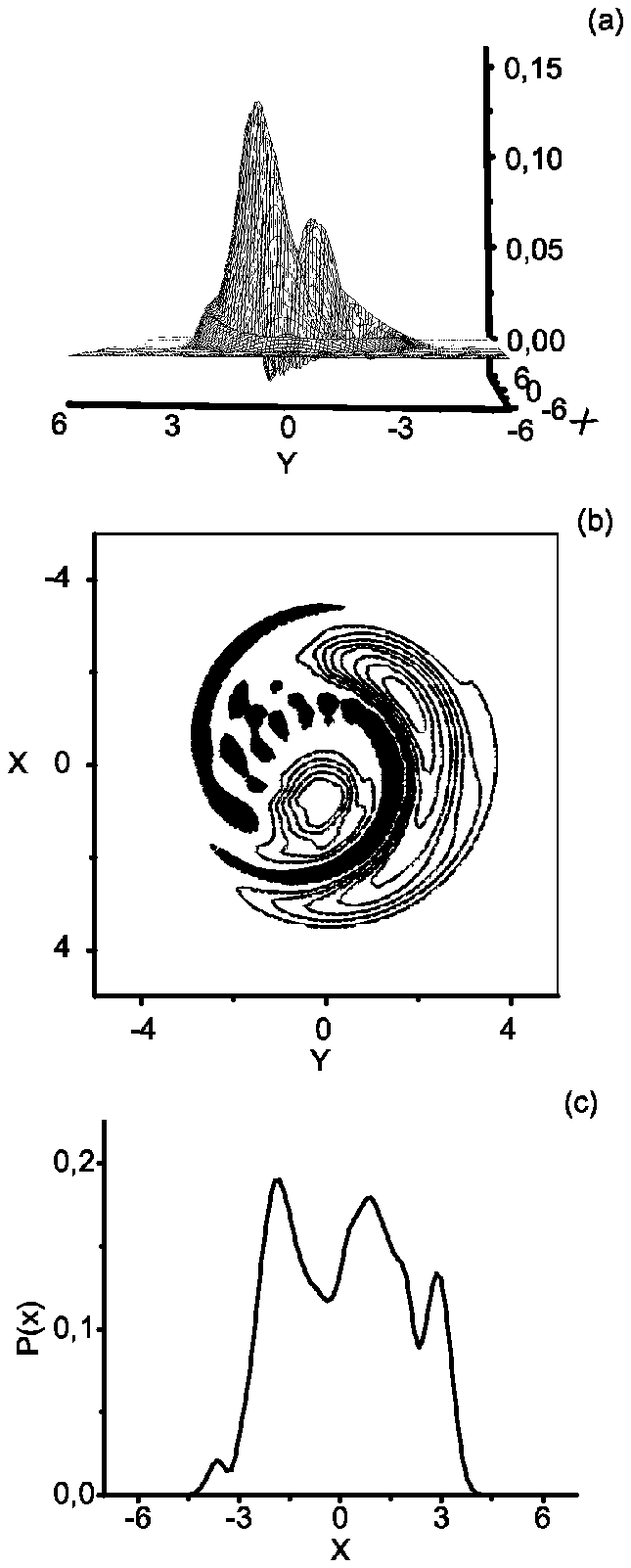}
\caption{The Wigner function (a); the contour plot of the Wigner
function (b); and the distribution of quadrature amplitude $P(x)$
for the driven dissipative AHO with time-modulation of the
nonlinearity. The parameters are in the range of bistability:
$\Delta/\gamma=-15, \chi(t)/\gamma=1.5(1+0.5\sin(\delta t)),
\delta/\gamma=5, f/\gamma=10.2$.}\label{scaled}
\end{figure}

\subsection{\label{sec: quant_inter_bis}Quantum interference pattern assisted by the bistability.}
First, we consider the case of time-modulated driving force. In
the limit $f_1 \ll f_0$ the system is reduced effectively to the
model of a single driven anharmonic oscillator, which exhibits
bistability for the definite range of the parameters: $\chi_0$,
$\Delta$, $f_0$, and $\gamma$ (see for example
\cite{duffing_force}, \cite{SR}, \cite{c}, \cite{drumm}), however,
for negative detuning, $\Delta < 0$. In this case, the
semiclassical steady-state oscillatory excitation number
$|\alpha|^{2}$, as solution of Eq. (\ref{eq:alpha_force}),
displays hysteresis behavior, while quantum mechanical mean
excitation number $\langle a^{+}a \rangle$ does not show any
hysteresis in the critical range due to quantum statistical
averaging \cite{I}. The analysis of the stochastic trajectories
for an expectation number $n_{\xi}(t)=\langle
\psi_{\xi}|a^{+}a|\psi_{\xi} \rangle$ shows that the system spends
most of its time close to one of the semiclassical bistable
solutions with quantum-interstate transitions, occurring at random
intervals \cite{SR}. On increasing the amplitude $f_1$ a new
channel of stimulated interstate transitions is raised between the
semiclassical bistable solutions. Their contribution for the
further increasing of $f_{1}$ at $f_{1} \leq f_{0}$ leads to the
emergence of a chaotic regime.

In another scenario of transition from regular regime to
bistability and chaos the modulation frequency $\Omega$ may be
varied, with the other parameters unchanged. In the range $\Omega
\ll \gamma$ the modulation of the system is adiabatic. On
increasing modulation frequency the system oscillates between the
two possible metastable states. At $\Omega \geq \gamma$ a strong
entanglement of these states occurs, and the system comes to a
chaotic regime. As can be seen from the numerical calculations, the
analogous results are also obtained from the case (i), i.e.,
for the case of modulation of the parameter $\chi (t)$. Some
differences between operational regimes of these two models will
be considered below.

The Wigner function is determined by the expression
\begin{equation} \label{eq: Wigner_alpha_rho}
W(\alpha, t)=\frac{2}{\pi^{2}}e^{-2|\alpha|^{2}}\int d^2\beta
\langle -\beta|\rho(t)|\beta\rangle
e^{-2(\beta\alpha^{*}-\beta^{*}\alpha)},
\end{equation}
which is obtained from the Eq. (\ref{eq:WigRT}), where, however,
the matrix elements of the density operator of the full
dissipative system are used,
 $\rho_{nm}(t)=M(\langle
n|\psi_{\xi}(t)\rangle \langle \psi_{\xi}(t)|m\rangle)$. The
coefficients $W_{mn}(r,\theta)$ are the Fourier transform of
matrix elements of the Wigner characteristic function (see Eq.
(\ref{eq:Wigf})).

As calculations show, the most striking signature of quantum
bistability in the presence of time-modulation is the appearance
of quantum-interference with negative regions in the Wigner
function. We illustrate quantum-interference pattern in phase
space and in the bistable operational regime on Figs. \ref{fig:bi}-\ref{fig:scaled}.
More importantly, we clearly observe (Fig. \ref{fig:bi}(a)) that the Wigner
function displays two peaks around the bistable semiclassical
solutions as well as the negative part between them that indicates
quantum-interference. The ranges of negativity are indicated in
details on Fig. \ref{fig:bi}(b). Another peculiarity is that the Wigner
function is nonstationary and the interference patterns take place
for the definite time intervals $\gamma t_{k}= 6.9 + \frac{2 \pi
k}{\delta}\gamma$ (k=0,1,2...), exceeding transient time, at which
the mean excitation number $\langle n\rangle $ reaches its maximal
value. The interference pattern is destroyed as the
time-modulation is decreased. Indeed, as it is shown on Fig. \ref{fig:bi}(c),
in the case $f_1=0$ the Wigner function is positive in all
phase-space and indicates quantum bistability in accordance with
the analytical results \cite{I}. Examples of the curves for the
ensemble-averaged excitation number distributions $P_{n}=\langle
n|\rho|n\rangle$ are demonstrated on Fig. \ref{fig:bi}(d) for the two cases
of modulated and non-modulated dynamics described on the Figs.
3(a) and 3(c). In correspondence with the bistable behavior of the
system we find a developing bimodal structure of $P_{n}=\langle
n|\rho|n\rangle$ for the oscillatory mode. However, this structure
is less pronounced for the case of non-modulated stationary
dynamics and is well resolved for the case of time-modulation of
the driving force, as we can also observe from the results for the
Wigner functions. The reason is that the interstate transitions
due to time-modulation of the driving force approximately
equalizes the populations of the states.

Note, that these results are given for a choice of parameters
which ensures that the oscillator is in the specific bistable
regime operating with small main excitation numbers. Nevertheless,
it is possible to find the parameters that would enable us to
observe quantum-interference pattern on a macroscopic level of
excitation numbers. For this goal, we use the scaling properties
of the driven AHO recently proposed in \cite{X1}. Indeed, it is
easy to verify that Eqs. (\ref{eq:alpha}) and
(\ref{eq:alpha_force}) are invariant with respect to the following
scaling transformation of the amplitude $\alpha\rightarrow
\alpha^{'}=\lambda\alpha$, where $\lambda$ is a real positive
dimensionless coefficient, if the oscillatory parameters are
correspondingly transformed as
$\Delta\rightarrow\Delta^{'}=\Delta+\chi(1-1/\lambda^{2})$, $
\chi\rightarrow\chi^{'}=\chi/\lambda^{2}$, $ f\rightarrow
f^{'}=\lambda f$, $\gamma\rightarrow \gamma^{'}=\gamma$. This
scaling properties of the semiclassical equations signifies that
for the definite different sets of the parameters the phase-space
trajectories have the same form and differ from each other only in
a scale. It has been numerically shown that this scaling parameter
is also approximately realized in a quantum ensemble theory in the
presence of quantum noise, for wider ranges of the parameters, but
not for large values of the parameter $\chi/\gamma$, i.e., in a
deeper quantum regime. We calculate the Wigner function and its
contour plot for the parameters obtained from the data of the Fig.
3(a) with the scaling parameter $\lambda=5$. The results presented
in Fig. \ref{fig:scaled}(a) and Fig. \ref{fig:scaled}(b), particularly, illustrate the quantum
interference pattern for a macroscopic level of the mean
excitation number, $\langle n\rangle=40$. As we see, the
negative part of the Wigner function decreases with increasing the
mean oscillatory excitation number. The result for time evolution
of the mean oscillatory excitation number averaged over quantum
trajectories is depicted in Fig. \ref{fig:scaled}(c). For such parameters the
system still operates in a bistable regime. We indicate the
hysteresis curve for the corresponding semiclassical solution for
the excitation number $|\alpha|^{2}$ on Fig. \ref{fig:scaled}(d) for the
stationary case: $f_{1} = 0$.

 Analogous results are obtained for the case (i), that is the time-modulation
of the strength of the nonlinearity. The results of numerical
calculations of the Wigner function and the corresponding
distribution of quadrature amplitude P(x) in the operational
regime of bistability are shown on Fig. \ref{fig:scaled}.

 Note that these interference effects have been demonstrated for the strong
quantum regimes of high nonlinearity. As a consequence, the oscillatory
excitation number is small in this bistable regime. Below we consider the
regime of chaotic dynamics where quantum-interference pattern is realized with
relatively high level of excitation numbers.
\begin{figure}
\includegraphics[width=8.6cm]{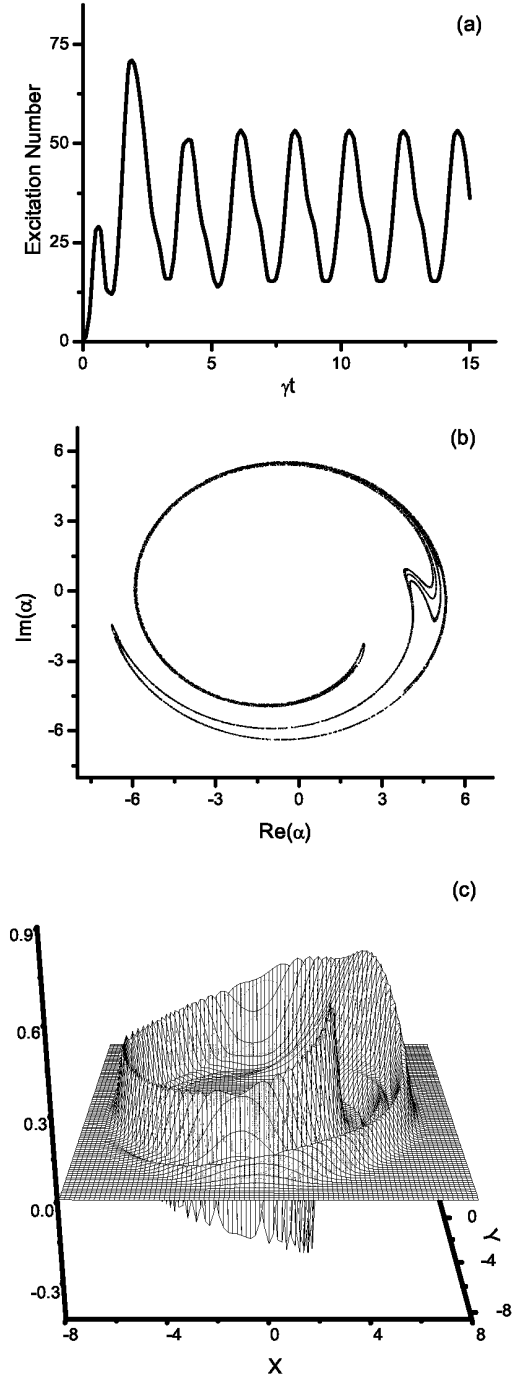}
\caption{(a) The mean excitation number; (b) the Poincar$\acute{e}$ section
($\approx 20000$ points) for the dimensionless
complex amplitude $\alpha$, plotted at times of the constant phase
$\delta t_{k}= 2 \pi k$ (k=0,1,2...), when the maximal
interference pattern on the Wigner function (c) is realized, for
the case of time-modulated nonlinearity $\chi(t)$. The parameters
are in the range of chaos: $\Delta/\gamma=-5,
\chi(t)/\gamma=0.2(1+0.75\sin(\delta t)), \delta/\gamma=3,
f/\gamma=10$.}\label{fig:chaos}
\end{figure}

\subsection{\label{sec:quant_inter_chaos}Negative Wigner functions for the chaotic dynamics.}
For both models of AHO corresponding to two forms of
time-modulations (cases (i) and (ii)), the dynamics of the system
is chaotic in the ranges: $\delta\geq\gamma$ or $\Omega\geq\gamma$
and $\chi_{0}\simeq \chi_{1}$ or $f_{0}\simeq f_{1}$ and for
negative detuning. Thus, the ways to realize the controlling
transition from the regular to chaotic dynamics through the
intermediate ranges of bistability are to vary the strength
$\chi_{1}$ or $f_{1}$ of the modulation processes in the ranges
from $\chi_{1}\ll\chi_{0}$ or $f_{1}\ll f_{0}$ to
$\chi_{1}\leq\chi_{0}$ or $f_{1}\leq f_{0}$. In order to get a
quantitative measure for chaotic dynamics, we have chosen the
analysis on the base of the $Poincar\acute{e}$ section.

We note, that the order-to-chaos transition for dissipative AHO
under time-modulated driving force was already analyzed \cite
{X1}. Therefore, here we concentrate on investigation of quantum
interference phenomenon considering AHO with time-modulated
nonlinearity. The results of the ensemble-averaged numerical
calculation of the mean excitation number, the Poincar$\acute{e}$
section and the Wigner function are shown on Figs. \ref{fig:chaos}(a)-\ref{fig:chaos}(c),
respectively. The mean excitation number of the driven AHO versus
dimensionless time is depicted in Fig. \ref{fig:chaos}(a). In contrast to the
semiclassical result its quantum ensemble counterpart (Fig. \ref{fig:chaos}(a))
has clear regular periodic behavior for time intervals exceeding
the characteristic dissipation time, due to ensemble averaging.
The Fig. \ref{fig:chaos}(b) clearly indicates the classical strong attractors
with fractal structure that are typical for a chaotic
$Poincar\acute{e}$ section. Thus, the Wigner function Fig. \ref{fig:chaos}(c)
reflects the chaotic dynamics, its contour plots in the (x,y)
plane are similar to the $Poincar\acute{e}$ section. However, the
Wigner functions have regions of negative values for the definite
time intervals. The example depicted on Fig. \ref{fig:chaos}(c) corresponds to
time intervals $\gamma t_{k}= 6 + \frac{2 \pi k}{\delta}\gamma$
(k=0,1,2...), for which the mean excitation number reaches a
macroscopic level, i.e., $n=52$.

\section{CONCLUSION}
We have numerically studied the phenomena at the overlap of
bistability, chaos and quantum-interference for the driven and
damped AHO with time-dependent parameters. We have pointed out,
that the time-modulation of the oscillatory parameters, that are
the strength of $\chi(3)$ nonlinearity or the amplitude of the
driving force, leads to formation of the quantum-interference
patterns in phase-space in over transient regimes, for the
definite time intervals exceeding the transient dissipation time
$t \gg \gamma^{-1}$. These effects are displayed as the negative
values of the Wigner functions in phase-space. Quantum
interference patterns take place for both bistable and chaotic
operational regimes and come from the oscillation between possible
metastable states of semiclassical dynamics due to
time-modulation. We have also demonstrated that the
time-modulation of the nonlinearity parameter for the regular
operational regime of AHO, essentially improves the degree of
sub-Poissonian statistics of oscillatory mean excitation number.
In this spirit, we emphasize that the idea of improving the degree
of quantum effects as well as obtaining qualitatively new quantum
effects due to appropriately time-modulation of open quantum
systems was recently exploited for formation of high degree
continuous variable entanglement in nondegenerate optical
parametric oscillator \cite{ENT}.

\begin{acknowledgments}
Acknowledgments: We acknowledge helpful discussions with S. B. Manvelyan, as well as financial
support from CRDF/NFSAT under Grant UCEP 07/02,
and ISTC under Grants A-1451 and A-1606.
\end{acknowledgments}

\end{document}